\documentclass[prb,aps,twocolumn,amssymb,amsmath,epsfig]{revtex4}
\newcommand{\VEC}[1]{{\mathbf{#1}}}
\newcommand{\TO}{,\ldots,} 
\newcommand{\Dh}{\ensuremath{\Delta^2}} 
\newcommand{\Dhm}{\ensuremath{\Delta^2}--model} 
\newcommand{\Dq}{\ensuremath{\Delta^4}} 
\newcommand{\Dqm}{\ensuremath{\Delta^4}--model} 
 
\newcommand{\Dgm}{\ensuremath{\Delta^{\gamma}}--model} 
\newcommand{\z}{\ensuremath{\zeta}} 
\newcommand{\mean}[1]{\langle #1 \rangle}
\usepackage{epsfig}

\begin{document}
\title{Depinning of elastic manifolds}
\author{Alberto Rosso, Alexander K. Hartmann\footnote{
present address: Institut f\"{u}r Theoretische Physik, Universit\"{a}t 
G\"{o}ttingen, Bunsenstr. 9, 37073 G\"{o}ttingen, Germany
}, Werner Krauth}
\email{rosso@lps.ens.fr; hartmann@theorie.physik.uni-goettingen.de; 
krauth@lps.ens.fr}

\affiliation{CNRS-Laboratoire de Physique Statistique \\
Ecole Normale Sup{\'{e}}rieure,\\
24, rue Lhomond, 75231 Paris Cedex 05, France}

\begin{abstract} 
We compute roughness exponents of elastic $d$--dimensional manifolds in
($d+1 $)--dimensional embedding spaces at the depinning transition for
$d=1 \TO 4$.  Our numerical method is rigorously based on a Hamiltonian
formulation; it allows  to determine the critical manifold in finite
samples for an arbitrary convex elastic energy.  For a harmonic elastic
energy (\Dhm), we find values of the roughness exponent between the
one-loop and the two-loop functional renormalization group result, in good
agreement with earlier cellular automata simulations.  We find that the
\Dhm\  is unstable with respect both to slight stiffening and to weakening
of the elastic potential.  Anharmonic corrections to the elastic energy
allow us to obtain the critical exponents of the quenched KPZ class.
\end{abstract}
\maketitle

Elastic manifolds in random media are an important issue of current
research in  statistical physics.\cite{Kardar,Barabasi} In the
zero-temperature motion of these manifolds, subject to a driving force
$f$, the ``depinning threshold'' $f_c$ plays a central role: For forces
$f > f_c$, the elastic manifold moves with finite velocity, while it
is pinned for $f \le f_c$.  Among the subjects studied at and around
$f_c$  are the ``creep" motion for $f < f_c$ at  finite temperature,
the scaling of the velocity for small forces $f \stackrel{>}{\sim} f_c$,
and the statistical properties of the pinned critical manifold at $f_c$,
especially its roughness exponent \z.

If one neglects velocity-dependent terms in the equations of motion
of the manifold, which one assumes to be a single-valued function
$h(\VEC{x},t)$ (no overhangs), its dynamics is governed by a functional
$E(\{h,\VEC{x}\})$ incorporating potential energy due to the driving
force $f$, the disorder $\eta(\VEC{x},h)$, as well as  its internal
elastic energy $E_{\text{el}}$,
\begin{equation}
\partial_t h(\VEC{x},t) =
- \frac{\partial E }{\partial h(\VEC{x}) } 
 = f + \eta(\VEC{x},h) - \frac{\partial E_{\text{el}}}{\partial h(\VEC{x})}.
\label{e:continuum_motion}
\end{equation}
Note that $\VEC{x}$ is a $d$--dimensional vector, in an embedding space
of dimension $d+1$. Dimensional analysis suggests that the harmonic
approximation for $E_{\text{el}}$ provides the only relevant term for 
the interplay between disorder and elasticity.  This yields
the quenched Edwards-Wilkinson (EW) equation
\begin{equation}
\partial_t h(\VEC{x},t) =
 f + \eta(\VEC{x},h) + a \nabla^2  h(\VEC{x}).
\label{e:EWquenched}
\end{equation} 

The model described by eq.~(\ref{e:EWquenched}) has been an important
testing ground for the concepts and techniques originally developed
in the field of  critical phenomena. Functional renormalization group
techniques\cite{Nattermann2,Narayan} were used to perturbatively compute
critical exponents of the quenched EW equation, which were believed to
describe generic driven manifolds at the depinning threshold.

According to the framework provided by the functional renormalization
group, the manifold  is flat ($\z = 0$) for $d \ge d_{\text{uc}} = 4$.
Below this upper critical dimension, the roughness exponent is expressed
in an $\epsilon$-expansion in $\epsilon =4-d$.  The first order (one-loop)
term \cite{Nattermann2,Narayan} of the expansion gives $\z=\epsilon/3$.
Initially, this result was believed exact\cite{Narayan} for all $d=1,2,3$.
However, Chauve {\em et al.} \cite{Chauve2} have obtained the two-loop
corrections, and found them to be non-zero.

\begin{figure}
\centerline{
\epsfig{figure=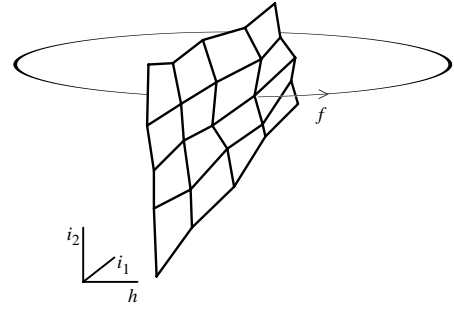,height=4.cm} }
\caption{ Schematic description of a $d=2$--dimensional manifold, driven
by a force density $f$. The ($d+1$)--dimensional embedding space is periodic
in the manifold's dimensions $i_1 \TO i_d$, but also in the variable $h$.
}
\label{f:2d_schema_periodic}
\end{figure}

In this work, we compute the roughness exponent of  critical manifolds
in finite ($d+1$)--dimensional samples for $d=1 \TO 4$.  Our approach
is rigorously based on a Hamiltonian formulation, as we study directly
the dynamics of eq.~(\ref{e:continuum_motion})  for a general convex
elastic energy $E_{\text{el}}$: A powerful numerical algorithm
allows to solve this problem for a slightly modified version of
eq.~(\ref{e:continuum_motion}), with a discretized  vector $\VEC{x}$
($\VEC{x} \rightarrow \VEC{i}$) and a continuous variable $h$.
The short-range elastic energy depends on the next-neighbor distances
$\Delta_{\VEC{i},\boldsymbol{\delta}}$ between  lattice point $\VEC{i}$
and its $2 d$ neighbors $\VEC{i}\pm \boldsymbol{\delta}_j$, as shown
in Fig.~\ref{f:2d_schema_periodic}:
\begin{equation}
 \frac{1}{2} \sum_{j=1}^{d}
\sum_{\pm \boldsymbol{\delta}_j} E_{\text{el}}
(|h_{\VEC{i}}- h_{\VEC{i} + \boldsymbol{\delta}_j}|).
\label{elastic_energy}
\end{equation}

We introduce periodic boundary conditions for $\VEC{x}$ and $h$
({\em cf} Fig.~\ref{f:2d_schema_periodic}).\cite{f-n-winding} Continuous,
periodic random potentials are constructed from $M$ normally
distributed  random variables.\cite{RossoKrauth3,RossoKrauth5} In
such a system, the sample-dependent critical force $f_c$ is well
defined: pinned  configuration with $v_{\VEC{i}}=0\ \forall \VEC{i}$
do (do not) exist for driving forces smaller (larger) than $f_c$.
Our algorithm\cite{RossoKrauth3} allows to decide quickly whether pinned
configurations exist at a given driving force, and then zooms in on the
critical force and the critical manifold.

For the first time, we are thus able to unambiguously compute the object
of most direct theoretical interest in this problem, namely the critical
manifold, in dimensions $d=2, 3, 4$.  We compute\cite{RossoKrauth2} the
roughness exponent \z\ from the disorder-averaged mean square deviations
of the critical manifolds $h^c$:
\begin{equation}
W^2(L) = \overline{
 \mean{(h^c -   \mean{ h^c})^2 } 
}
\sim L^{2\z}\ 
\text{for}\ L \rightarrow \infty.
\label{e:elongdef}
\end{equation}
As the width  of the manifold is of order $ \sim L^{\z}$, we
scale the lateral extension of the sample at least as $M \sim L^{\z}$.

Especially in higher dimensions, direct simulations of
eq.~(\ref{e:EWquenched}) are not viable, because they violate a
crucial no-passing theorem,\cite{Middleton} which we respect.

Because of the expected universality and the difficulties of direct
simulation, attempts to compute the critical exponents of the
quenched EW equation considered discrete systems with non-hamiltonian
dynamic rules.
These cellular automata models
\cite{Sneppen,Buldyrev} are not clearly connected with an equation of
motion in the continuum limit \cite{Vannimenus} and have often given only
rough estimates for \z.

Nevertheless, simulations using cellular automata illustrated
that critical elastic manifolds in random media are not generally
described by the quenched EW equation, but may fall into two broad
classes: The automata \cite{Nattermann2,Leschhorn} summarized in
Ref. \onlinecite{Nattermann1}, yield critical exponents which are close
to those obtained by  analytical work on eq.~(\ref{e:EWquenched}); on
the other hand the automata \cite{Sneppen,Buldyrev} summarized in Ref.
\onlinecite{BuldyrevII} give very different values for the exponents,
incompatible with the results obtained by renormalization group results,
but closer to experiment.\cite{Buldyrev,BuldyrevIII,Krim} There has been
much confusion about the actual values of these exponents for $d>1$. It
is not  clear whether the upper critical dimension of this class is
also  $d_{\text{uc}}=4$.

In our controlled Hamiltonian approach, we recover this very rich
behavior as a dependence of the roughness exponent \z\ on the functional
form of the elastic energy: We investigate elastic energies of the
form $E_{\text{el}} \propto |\Delta|^{\gamma}$, and refer to them
as $\Delta^{\gamma}$--models.  The \Dhm\ studies directly the
eq.~(\ref{e:EWquenched}).  Models with $\gamma  \neq 2$ differ in their
critical behavior from the harmonic model. A special role is played by
the \Dqm, which represents the first non-harmonic corrections of
a general elastic energy $E_{\text{el}} \sim  a \Dh/2 + b \Dq/12 + \ldots$

\begin{figure}
\centerline{ \epsfig{figure=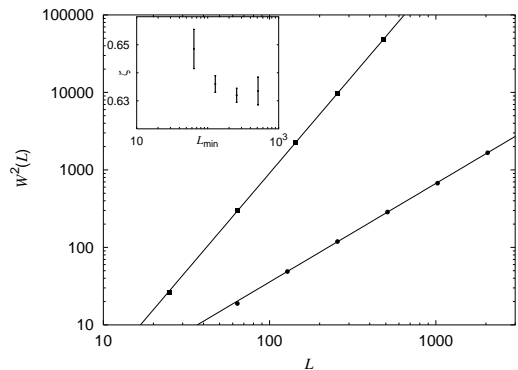,height=5.cm} }
\caption{Mean square extension $W^2$ vs system size for the
($1+1$)--dimensional elastic string. Upper curve: harmonic elastic
energy $E_{\text{el}}(\Delta)= \Dh/2$; $M \sim L^{1.5}$.
 Lower curve: elastic energy
$E_{\text{el}}(\Delta)= \Dq/4$; $M =  L$  . 
The inset shows the estimated value
of \z\ for all the data with $L >  L_{\min}$ as a function
of $L_{\min}$ (\Dqm). The absence of systematic trends leads us to conclude that
$\z_{\Dh} = 1.26 \pm 0.01$,  
$\z_{\Dq} = 0.635 \pm 0.005$.}
\label{f:1d_synopsis_with_inset} 
\end{figure}

We first discuss this important issue for an elastic string ($d=1$)
in a $2$--dimensional medium, which has been abundantly studied in the
past.  As shown in Fig.~\ref{f:1d_synopsis_with_inset}, we find for the
\Dhm\ a value of the roughness exponent $\z_{\Dh}=1.26 \pm 0.01$,
which compares well with the two-loop calculation of Chauve {\em et al.},
while clearly excluding the one-loop result $\z^{\text{oneloop}}_{\Dh}=1$.
A value in excess of one for the $(1+1)$--dimensional roughness has been found
in direct numerical integrations of the equations of motion,\cite{Jensen}
in Monte Carlo simulations,\cite{RossoKrauth1,RossoKrauth2} and in the
cellular automata of Ref. \onlinecite{Nattermann1}.

Figure~\ref{f:1d_synopsis_with_inset} also shows our results for
the \Dqm. We find \cite{f-n-inset} $\z_{\Dq}=0.635 \pm 0.005$.  This
exponent coincides with the value found in Ref. \onlinecite{BuldyrevII}.
In lattice models, we already showed that the roughness exponent does not
change for the $\Delta^6$-model and even in the presence of a  metric
constraint (bounded $|\Delta|$). In this sense, the \Dqm\ is only
one  representative of systems with a stronger than harmonic elastic
energy. The values obtained from Fig.~\ref{f:1d_synopsis_with_inset}
are much more precise than the earlier ones.

In this problem, the study of rotational invariance of the equations
of motion has occupied a crucial role. 
In fact, eq.~(\ref{e:EWquenched}), and therefore the equation of motion of 
our \Dhm,  is invariant under a tilt of the manifold
\begin{equation}
h(\VEC{x},t)  \rightarrow  h(\VEC{x},t) + \sum m_i\ x_i.
\label{e:tilt}
\end{equation}
It was observed \cite{Amaral} that the automata of
Ref. \onlinecite{BuldyrevII} present strong dependence of the velocity and
of the  value of $f_c$ on the parameters $m_i$.  In our \Dqm,
rotational invariance is manifestly broken as an elastic energy $ b
\Dq/12$ generates a piece
\begin{equation}
     b\ {\nabla}^2 h (\nabla h)^2
\label{e:quartic_motion}
\end{equation}
in the equation of motion.  From dimensional
analysis,\cite{Kardar,KardarAniso} among the orientation-dependent
differential operators, the KPZ term  $\sim (\nabla h)^2$ is more relevant
than the term eq.~(\ref{e:quartic_motion}).  The non-linear KPZ term
appears in  domain-growth models without disorder, where it is coupled
to the velocity.\cite{KPZ} Adding it to eq.~(\ref{e:EWquenched})
constitutes the quenched KPZ equation. Its critical exponents have not
been computed within the functional renormalization group,  and it is
not even clear how this term is generated in the absence of a finite
velocity, at depinning.  \cite{Narayan}

Kardar {\em et al.} \cite{Kardar,KardarAniso} have suggested the force
anisotropy due to the disorder as a possible generating mechanism,
but none of the automata models present such an anisotropy. However,
in their construction rules, a metric constraint is hidden. This
implements a strong elastic potential, and naturally generates terms
like eq.~(\ref{e:quartic_motion}). In our opinion, the main unresolved
theoretical issue is to understand how this term reduces to the KPZ term
under coarse-graining.

As discussed in Ref. \onlinecite{RossoKrauth2},  there is an easy way 
to  understand why the ($1+1$)--dimensional  \Dhm\
is unstable to higher-order corrections: A roughness $\z > 1$
implies\cite{TangLeschhornDongcomm} that the mean local elongation
$\mean{|\Delta|}$ grows with the system size at least as $L^{\z -1}$,
and thus diverges in the thermodynamic limit $L,M \rightarrow \infty$.
Higher-order terms in the elastic energy are thus trivially relevant in
one dimension.\cite{RossoKrauth2}

\begin{figure}
\centerline{ \epsfig{figure=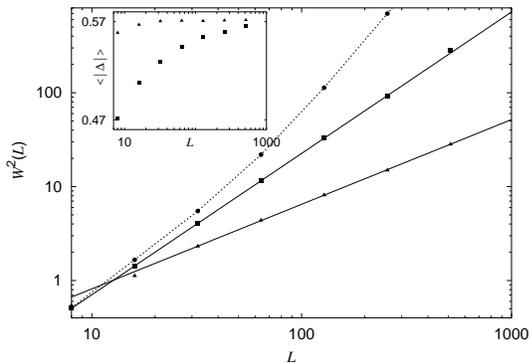,height=5.cm} }
\caption{ 
Mean square extension $W^2$ vs system size for the ($2+1$)--dimensional
elastic manifold for energies $E_{\text{el}}(\Delta)=\frac{1}{\gamma}
\Delta^{\gamma}$ for $\gamma = 1.8$ (upper, circles), $\gamma = 2$
(middle, squares) and $\gamma = 4$ anharmonic energy (lower, triangles);
$M=L$.  The inset shows the mean local elongation $\mean{|\Delta|}$ for
$\gamma=2$ and $\gamma=4$.  For $\gamma=1.8$, $\mean{|\Delta|}$
clearly diverges (not shown). Error bars are smaller than the symbol sizes.
}
\label{f:2d_synopsis_with_inset}
\end{figure}

In $2+1$ dimensions, we have studied  samples up to size
$L^2 \times M= 512^2 \times 512$.  For the \Dhm\ (shown in
Fig.~\ref{f:2d_synopsis_with_inset}), we find excellent scaling for
a roughness $\z_{\Dh}=0.753 \pm 0.002$, which again falls between
the results of the one-loop and two-loop renormalization group
calculations. Our result improves  by an order of magnitude the precision
of the  previous estimate.\cite{Nattermann1} Note that our algorithm
allows us to know, without invoking universality arguments,  that the
exponent $\z_{\Dh} \sim 0.753$ is that of the two-dimensional quenched
EW model.

In Fig.~\ref{f:2d_synopsis_with_inset}, we also show our results for the
\Dqm. We find a different value for \z, namely $\z_{\Dq}=0.45 \pm 0.01$.
This last value is  significantly smaller than  $\z \sim 0.48$ found in
Ref. \onlinecite{BuldyrevII} using a cellular automaton.

The two-dimensional \Dhm\ is not only unstable with respect to stronger
elastic potentials, but also to any weaker  ones.  We have  studied the
$\Delta^{\gamma}$-model, for $\gamma$ slightly below $2$: Already for
$\gamma = 1.8$ ({\em cf} Fig.~\ref{f:2d_synopsis_with_inset}), we find that the
roughness exponent changes drastically, as we obtain clear indications
that $\z_{\Delta^{1.8}} > 1$.  This implies that the surface breaks.

We also studied elastic energies $ \sim \sqrt{1 + \Dh}$ in the
($2+1$)--dimensional problem. In one dimension, this functional
form corresponds to the length of the string, and the  \Dhm\ is often
understood as the first term from an expansion of the length in powers of
$\Delta$. The square root is softer than its first expansion coefficient;
we again find values of $\z_{\sqrt{1 + \Dh}}$ in excess of one for large
system sizes.

As in $1+1$ dimensions, we also find for the ($2+1$)--dimensional \Dhm\
indications of a (much slower) divergence of the mean local extension,
as shown in Fig.~\ref{f:2d_synopsis_with_inset}. For the \Dhm, it seems
to diverge logarithmically, whereas for the \Dqm\ it saturates already
for very small systems. Again it appears that only the \Dqm\ has a proper
thermodynamic limit for the critical manifold.

\begin{figure}
\centerline{ \epsfig{figure=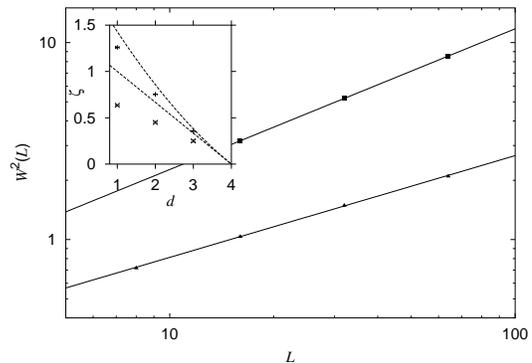,height=5.cm} }
\caption{ Mean square extension $W^2$ vs system size ($M=L$)
for the ($3+1$)--dimensional elastic manifold for energies
$E_{\text{el}}(\Delta)=\frac{1}{4\gamma}\Delta^{\gamma}$ for $\gamma = 2$
(harmonic energy, upper curve) and $\gamma = 4$ (lower curve).
The inset shows $\z_{\Dh}$ and $\z_{\Dq}$ vs $d$. The lines are the
one--loop and two--loop results for the \Dhm. 
}
\label{f:3d_synopsis}
\end{figure}

We also performed extensive computations in  $3+1$ and in $4+1$
dimensions, where we have reached sample sizes of $L^3 \times M=
64^3\times 64$ and $L^4 \times M = 32^4 \times 32$, respectively. For
the harmonic model in $3+1$ dimensions, we  obtain $\z_{\Dh} =0.355
\pm 0.01$ again in good  agreement with the numerical results of
Ref. \onlinecite{Leschhorn}, while our result $\z_{\Dh} =0.25 \pm 0.02$
for the \Dqm\ is in contradiction to Ref. \onlinecite{BuldyrevII} which
suggested an exponent $\z \sim 0.38$ larger than for  the harmonic case.
We checked very carefully that our estimate for \z\ is independent of
the parameters $a$ and $b$.

At last, our algorithm is able to investigate $4+1$
dimensions.\cite{RossoKrauthVannimenus} We have  computed manifolds
of sizes $L = 8, 16, 32 \, (L=M)$.  Even though we have only three
points, we find for the \Dhm\ an inconsistent fit to a functional form
$W^2(L) \sim L^{2\z}$ for sizes $L$ where it was already good in $3+1$
dimensions. At the upper critical dimension, one would expect logarithmic
behavior for $W^2(L)$. The most natural explanation of our findings
in $4+1$ dimensions is that the \Dhm's upper critical dimension is
$d_{\text{uc}}=4$.  This will still have to be confirmed for larger
sample sizes.

For the \Dqm, we also do not find a good fit for \z, but the deviations
from the functional form  $W^2(L) \sim L^{2\z}$ are less striking than
for the \Dhm. It is unclear to us whether the roughness exponent of the
\Dqm\  \emph{can}  be higher than for the \Dhm.

\begin{table}
\begin{center}
\begin{tabular}{|c|c|c|c|c|}
 \hline
$d$  & \multicolumn{2}{c|}{analytic} & $ \z_{\Dh}$  & $ \z_{\Dq}$ \\
\cline{2-3}
    & one-loop& two-loop & & \\ 
     1  & 1 & 1.44 &  1.26 $\pm$ 0.01  & 0.635 $\pm$ 0.005\\
\hline
    2 & 2/3 &0.86 & 0.753 $\pm$ 0.002  & 0.45 $\pm$ 0.01\\
\hline
   3 & 1/3&0.38 & 0.355 $\pm$ 0.01  & 0.25 $\pm$ 0.02 \\
 \hline 
 
 \end{tabular}
\end{center}
\caption{Roughness exponents as a function of dimension $d$ for the 
($d+1$)--dimensional driven manifold problem, both for the \Dhm\ 
(harmonic elastic energy) and the \Dqm.
\label{tabzeta2} }
\end{table}

In conclusion, we have directly computed roughness exponents for
$d$--dimensional manifolds in a ($d+1$)--dimensional embedding space for
$d=1 \TO 4$. Some of our findings nicely fit into the existing theoretical
framework. For example, the upper critical dimension $d_{\text{uc}}=4$,
conjectured on the basis of dimensional power counting, is consistent
with our data for the \Dhm.

In the harmonic case, agreement with the two-loop calculation is fair,
and the exactness of the one-loop result can be excluded ({\em cf}
inset of Fig.~\ref{f:3d_synopsis}).  Our numerical work relies on no
additional hypothesis, as we compute well-defined critical manifolds
in finite systems. Even for the harmonic case, though, we would like to
understand the divergence of the local scale, for which we find evidence
even in $d =2$: slightly sub-harmonic potentials lead to a break-up of
the critical surface.

For the first time we have presented a Hamiltonian model which reproduces
the results of the quenched KPZ class.  The connection between our \Dgm\
and the  equations of motion in the continuum limit is transparent.
The analysis of the \Dqm\ gives a very precise estimate for the value of
\z\ as a function of dimension $d$.  It would be very interesting to
understand how the local differential operators, as $ {\nabla}^2 h
(\nabla h)^2$, which correspond to elastic Hamiltonians, 
renormalize into more relevant terms, as the KPZ term.

Acknowledgments: We thank   P. Le Doussal for very  helpful discussions
all along this work, and also acknowledge stimulating discussions with
D.~S.~Fisher and K.~J.~Wiese.  Our computations  were performed on
clusters of workstations at Ecole Normale Sup\'erieure, the Institut
f\"ur Theoretische Physik of the Universit\"at Magdeburg (Germany)
and at the Paderborn Center for Parallel Computing (Germany). AKH was
financially supported by the DFG (Deutsche Forschungsgemeinschaft)
under grant Ha 3169/1-1.


\begin{thebibliography}{10}
\bibitem{f-n-inset} The relation $W^2 \sim L^{2 \z}$ holds only in the
limit of large $L$. Dependence of the fit on the smallest of the
included $L$-values is checked in the inset of
Fig.~\ref{f:1d_synopsis_with_inset}, which testifies to  the absence of
systematic trends for  $L_{\min} > 16$.
\end{thebibliography}
\end{document}